\documentclass[a4paper,fleqn,usenatbib]{mnras}

\usepackage{graphicx}	
\usepackage{multirow}
\usepackage{hyperref}

\usepackage{times}
\usepackage{amssymb,amsmath}

\usepackage{color}
\def\si{\color{black}}

\def\lsim{ \lower .75ex\hbox{$\sim$} \llap{\raise .27ex \hbox{$<$}} }
\def\gsim{ \lower .75ex \hbox{$\sim$} \llap{\raise .27ex \hbox{$>$}} }

\title[RIAFs and BL Lac neutrino emission] 
{Neutrino emission from BL Lac objects: the role of radiatively inefficient accretion flows
}

\author[Righi, Tavecchio \& Inoue]
{C. Righi$^{1,2,3}$
, F. Tavecchio$^2$, S. Inoue$^4$ \\
$^1$ Universit{\`a} degli Studi dell'Insubria, Via Valleggio 11, I-22100 Como, Italy\\
$^2$ INAF -- Osservatorio Astronomico di Brera, via E. Bianchi 46, I--23807 Merate, Italy\\
$^3$ INFN -- Sezione di Genova, Via Dodecaneso 33, I-16146 Genova, Italy\\
$^4$ Astrophysical Big Bang Laboratory, RIKEN, 2-1 Hirosawa, Wako 351-0198 Japan
}
\begin{document}



\maketitle

\begin{abstract}
{\si
The origin of the astrophysical high-energy neutrinos discovered by IceCube is currently a major mystery. The recent detection of IceCube-170922A, a $\sim$300 TeV neutrino potentially correlated with the flaring $\gamma$-ray source TXS 0506+056, directs attention toward BL Lac objects (BL Lacs), the subclass of blazars with weak emission lines. While high-energy neutrinos can be produced via photohadronic interactions between protons accelerated in their jets and ambient low-energy photons, the density of the latter in such objects had generally been thought to be too low for efficient neutrino emission. Here we consider the role of radiatively inefficient accretion flows (RIAFs), which can plausibly exist in the nuclei of BL Lacs, as the source of target photons for neutrino production. Based on simple model prescriptions for the spectra of RIAFs at different accretion rates, we find that they can be sufficienly intense to allow appreciable neutrino emission for the class of low-synchrotron-peak BL Lacs such as TXS 0506+056. In constrast, for high-synchrotron-peak BL Lacs including Mkn 421 and Mkn 501, the contribution of RIAFs is subdominant and their neutrino production efficiency can remain low, consistent with their non-detection by IceCube to date.
}
\end{abstract}


\section{Introduction}
{\si
The origin of the astrophysical neutrinos with energies above 100 TeV detected by IceCube (Aartsen et al. 2013) is  currently mysterious. The main observational challenges are their limited detection rates ($\sim 60$ events since 2010) and large localization uncertainties ($\sim1^\circ$ for muon track events and larger for cascade events). The observed isotropy of their distribution in the sky suggests a predominantly extragalactic origin.}

{\si
The most likely production mechanism of high-energy neutrinos in astrophysical environments is the acceleration of protons to sufficiently high energies $E_p$, followed by their inelastic collisions with ambient gas or low-energy photons. Such interactions generate charged pions that subsequently decay into secondary particles including neutrinos, with typical energy $E_\nu \sim 0.05 E_p$ (e.g. Dermer \& Menon 2009).
}
{\si
Among the various astrophysical sources that have been proposed (for reviews, see e.g. Ahlers \& Halzen 2015; Meszaros 2017), one of the most promising are blazars. Recognized as active galactic nuclei with relativistic jets oriented nearly towards the observer, the spectral energy distributions (SEDs) of their luminous, variable and broadband non-thermal emission
are typically characterized by two humps (Madejski \& Sikora 2016). The first one peaking in the infrared to soft X-ray band is understood as synchrotron emission of electrons accelerated inside the jet. The second one peaking in the $\gamma$-ray band is often interpreted as inverse Compton (IC) upscattering of ambient low-energy photons by the same electrons. It is plausible that protons are accelerated in the same conditions up to ultra-high energies, which can undergo $p\gamma$ interactions with ambient photons to produce high-energy neutrinos (e.g. Mannheim 1995). Neutrinos with $E_\nu \sim 300$ TeV require interactions between protons with $E_p \ge 6$ PeV and photons with energies above the photopion threshold, $\epsilon \ge m_\pi m_p c^4 / E_p \approx 10^2-10^3$eV, in the UV to soft X-ray range. In some models, the secondary cascade emission triggered by the same $p\gamma$ interactions can dominate the $\gamma$-rays (e.g. Mannheim 1993).}

{\si
Blazars can be categorised into two main subclasses: flat spectrum radio quasars (FSRQs) and BL Lac objects (BL Lacs). FSRQs are relatively more powerful, especially in GeV $\gamma$-rays, and exhibit strong emission lines. The latter is a signature of intense optical-UV photons from the broad line region outside the jet, likely photoexcited by a radiatively efficient accretion disk around the supermassive black hole (SMBH) at the nucleus (e.g. Ghisellini et al. 2010). IC upscattering of such external photons impinging into the jet, referred to as external Compton (EC) emission, can dominate the $\gamma$-rays. As the same photons can also serve as effective targets for $p\gamma$ interactions, FSRQs have been considered promising neutrino emitters (e.g. Murase et al. 2014; Kadler et al. 2016). However, a dominant contribution of FSRQs to the diffuse neutrino flux observed by IceCube is disfavored, since their low surface density and high luminosity is at odds with upper limits on source clustering (Kowalski 2015) or multiplet events (Murase \& Waxman 2016).
}
{\si
BL Lacs are relatively less powerful, and display weak or no emission lines, indicating the lack of strong external radiation fields. Their $\gamma$-ray luminosity is comparable to the synchrotron luminosity, but extends to higher energies. BL Lacs can be further subdivided depending on the peak energy of their SED components, with low-synchrotron-peak BL Lacs (LBLs) emitting up to hundreds of GeV, and high-synchrotron-peak BL Lacs (HBLs) up to tens of TeV (Ackermann et al. 2015)\footnote{Note that the abbreviations here differ from ``HSP'', ``ISP'' and ``LSP'' used in Ackermann et al. (2015).}. The $\gamma$-rays observed in BL Lacs can generally be well explained as synchrotron self-Compton (SSC) emission, i.e. IC emission by electrons accelerated in the jet upscattering their own synchrotron emission. Given their low power and inferred weak radiation fields, the neutrino production efficiency for BL Lacs have often been thought to be low (Murase et al. 2014; see however, Sec. 4 and 5).
}

{\si 
The recent finding that the likely counterpart of the $\sim$300 TeV neutrino IceCube-170922A is TXS 0506+056, a BL Lac (Aartsen et al. 2018), is therefore not trivial to interpret. Note that TXS 0506+056 is likely an LBL or possibly an intermediate-synchrotron-peak BL Lac (IBL; see below). The picture is further complicated by the fact that the HBLs Mkn 421 and Mkn 501 are still undetected in high-energy neutrinos, despite being more prominent $\gamma$-ray emitters (Aartsen et al. 2017).

In this context, a potential source of external photons for BL Lacs that has hardly been discussed in the literature is 
radiatively inefficient accretion flows (RIAFs). It is quite plausible that the nuclei of BL Lacs host RIAFs (e.g. Ghisellini, Maraschi \& Tavecchio 2009), which are expected when the mass accretion rate $\dot{M}$ onto the central SMBH is lower than a critical value (Narayan \& Yi 1995; Yuan \& Narayan 2014).
Notwithstanding their lower emissivity compared to standard accretion disks, the spectra of RIAFs are expected to span a broader frequency range, and vary non-trivially with $\dot{M}$. This study focuses on the role of RIAFs as external target photons for $p\gamma$ neutrino production in BL Lacs, which can have various interesting implications, including marked differences between LBLs and HBLs.
}

\section{BL Lac spectral energy distributions}
\label{sec:SED}

{\si
Using a sample of 747 blazars (299 BL Lacs and 448 FSRQs) detected by \textit{Fermi}-LAT with known redshifts from the 3LAC catalog (Ackermann et al. 2015), Ghisellini et al. 2017 (hereafter G17) confirmed the evidence for a spectral sequence, a systematic trend among the SEDs of all blazars that had been found in previous studies (Fossati et al. 1998). To parameterise their average SEDs, G17 used a phenomenological function consisting of two broken power laws connecting with a power law describing the radio emission. They found that with increasing luminosity, BL Lacs have lower peak frequency, softer $\gamma$-ray slope and larger dominance of the high-energy component.
}

\begin{table*} 
\centering
\begin{tabular}{cllllllllllll }
\hline
\hline
& $\log <L_{\rm bol}>$  &$\alpha_1$ &$\alpha_2$  &$\alpha_3$ &$\nu_{\rm t}$   &$\nu_{\rm S}$ &$\nu_{\rm C}$ &$\nu_{\rm cut, S}$
&$\nu_{\rm cut, C}$ &$\nu_{\rm S}L(\nu_{\rm S})$ &CD  &N\\
& $\log$ (erg s$^{-1}$)    &           &            &              &Hz            &Hz           &Hz       &Hz  &Hz  &erg s$^{-1}$  & \\
\hline   
{\bf LBL} &$47.2$   &0.65  &1.3 & 0.62  & 3e11 & 1e12 & 3e21 & 5e16 &7e26   & 1e46 & 1   &71  \\

\hline   
{\bf IBL} &$46.2$   &0.7  &1.3  &0.8  &2.5e11 &5e14 &1e23 &6e18 &8e26   &8e44 &0.7  &21 \\

\hline   
{\bf HBL} & $45.8$ &0.68  &1.2  &0.8  &1e11 &9e16 &5e24   &4e19 &5e27 &4e44   &0.4  &18  \\
\hline
\hline
\end{tabular}
\caption{
\si{Parameters for the phenomenological SEDs plotted in Fig. 1, where $\alpha_{\rm R}=-0.1$ was fixed.
See G17 for detailed definition of the parameters.}
}
\label{para}
\end{table*}

%
%

{\si
Ackermann et al. (2015) proposed a subclassification of BL Lacs, based on the peak frequency $\nu_S$ of the synchrotron SED component: LBLs with $\nu_S<10^{14}$ Hz,
IBLs with $10^{14}$ Hz $<\nu_S<10^{15}$ Hz, and
HBLs with $\nu_S>10^{15}$ Hz.
This ``{\it Fermi}" classification scheme can be applied to 110 out of the 299 BL Lacs in G17 with sufficient data, resulting in 71 LBLs, 21 IBLs and 18 HBLs.
}


\begin{figure}
\hspace{-0.6cm}
\centering
  \includegraphics[width=0.46\textwidth]{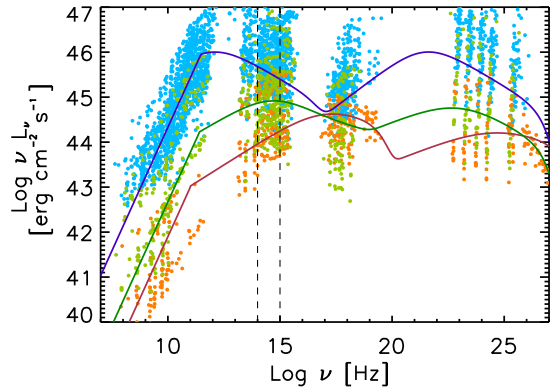}
  \caption{{\si Observed SEDs (filled circles) compared with average parameterized models for the three subclasses of BL Lacs: LBL (top, blue), IBL (middle, green), HBL (bottom, orange).}}
  \label{sedtot}
\end{figure}

For the purpose of estimating typical values of the observed bolometric luminosity, intrinsic radiative power, and jet power for each BL Lac subclass, we parameterize the average SEDs of their non-thermal emission, using the phenomenological model of G17 and assuming $\alpha_R=-0.1$ for the radio spectral index. For each subclass, the model is compared with the data in Fig.\ref{sedtot}, and the ten model parameters that were determined are listed in Tab. \ref{para}. From the model, it is straightforward to evaluate the average, isotropic-equivalent bolometric luminosity $L_{\rm bol}$. Assuming a jet bulk Lorenz factor $\Gamma_j=15$ (Ghisellini et al. 2010), the beaming-corrected power in radiation can be estimated as $P_{\rm rad}=L_{\rm bol}/\Gamma_j^2$ (e.g. Celotti \& Ghisellini 2008), with values listed in Table \ref{tab:adaf}. LBLs approximately have $L_{\rm bol}$ ten times larger than IBLs, and 25 times larger than HBLs. We note that our aim here is not detailed spectral modelling of these SEDs.



\section{{\si Radiatively inefficient accretion flows}}
\label{sec:RIAF}

A key quantity that regulates the properties of the accretion flow onto the SMBH with mass $M_{\rm BH}$ is the mass accretion rate in Eddington units $\dot{m}\equiv\dot{M}/\dot{M}_{\rm Edd}$, where $\dot{M}_{\rm Edd} \equiv L_{\rm Edd}/\eta_{\rm acc} c^2$, $L_{\rm Edd}=4\pi G M_{\rm BH} m_p c / \sigma_T$, and $\eta_{\rm acc}=0.1$ is a nominal accretion efficiency. When $\dot{m} \gtrsim 10^{-2}$, a standard accretion disk is expected that is geometrically thin, optically thick, and radiatively efficient (Shakura \& Sunyaev 1973). In contrast, when $\dot{m}\lesssim10^{-2}$, a transition to RIAFs is supported by both theory and observations (Yuan \& Narayan 2014). In RIAFs, the density of the accretion flow is low enough that thermal protons cannot transfer their energy effectively to thermal electrons via collisional processes. As a consequence, the proton temperature remains close to the virial value, much hotter than in standard accretion disks, and the flow becomes geometrically thick and optically thin while radiating inefficiently. As opposed to standard accretion disks with predominantly thermal spectra in the optical-UV range, the spectra of RIAFs can be more broadband and complex, comprising multiple components spanning the radio to soft $\gamma$-ray bands.


The trends in the observed phenomenology of FSRQs and BL Lacs have been adequately interpreted in terms of a sequence in the total power (energy flux) in the jet $P_{\rm jet}$ correlated with $\dot{M}$, together with a transition from standard accretion disks in the nuclei of FSRQs to RIAFs in those of BL Lacs  (e.g. Ghisellini et al. 2009). Within the BL Lac population, the trends among LBLs, IBLs and HBLs may also be understood as a sequence in $P_{\rm jet}$ and $\dot{M}$ . Here we discuss the emission expected from RIAFs for each BL Lac subclass.

First, we adopt a simple scaling between $P_{\rm jet}$ and $\dot{M}$ (e.g. Ghisellini et al. 2010),
\begin{equation}
P_{\rm jet} \approx \eta_j \dot{M} c^2,
\end{equation}
where a value $\eta_j \sim 1$ is supported for the jet formation efficiency through modeling of {\it Fermi}-LAT blazars (Ghisellini et al. 2014) as well as numerical simulations of magnetically-driven jet formation (Tchekhovskoy et al. 2011). On the other hand, a relation on average between $P_{\rm jet}$ and $P_{\rm rad}$ as derived from the SEDs in Sec. 2 is indicated by several studies,
\begin{equation}
P_{\rm jet} \approx P_{\rm rad}/\eta_{\rm rad},
\end{equation}
with $\eta_{\rm rad} \sim 0.1$ (Celotti \& Ghisellini 2008, Nemmen et al. 2012).
Assuming for simplicity fixed values of $M_{\rm BH}=10^{9} M_{\sun}$, $\dot{m}$ can be estimated for each BL Lac subclass from their average $P_{\rm rad}$ as
\begin{equation}
\dot{m} \approx \frac{\eta_{\rm acc}}{\eta_j}\frac{P_{\rm jet}}{L_{\rm Edd}} \approx \frac{\eta_{\rm acc}}{\eta_j \eta_{\rm rad}}\frac{P_{\rm rad}}{L_{\rm Edd}},
\end{equation}
resulting in $\dot{m} = 5.7\times 10^{-3}$, $5 \times 10^{-4}$ and $3\times 10^{-4}$ for LBLs, IBLs and LBLs, respectively.

The details of the emission from RIAFs can be model-dependent (Yuan \& Narayan 2014). For concreteness,
we adopt the advection-dominated accretion flow (ADAF) model of Mahadevan (1997), who provide simple prescriptions for calculating the spectra for different parameters. We assume viscosity parameter $\alpha=0.3$, ratio of gas pressure to total pressure $\beta=0.5$, minimum radius $r_{\rm min}=3 r_S$ and maximum radius $r_{\rm max}=10^3 r_S$ in units of $r_S = 2 G M_{\rm BH}/c^2$ (for more details, see Mahadevan 1997).

\begin{figure}
\hspace{-0.truecm}
  \includegraphics[width=.45\textwidth]{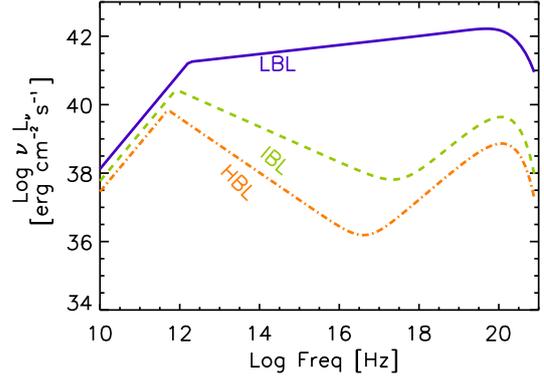} 
  \caption{
  {\si RIAF spectra expected for the three subclasses of BL Lacs: LBL (solid blue), IBL (dashed green) and HBL (dot-dashed orange).}}
  \label{fig:adafspectrum}
\end{figure}

For the three subclasses of BL Lacs, Fig. \ref{fig:adafspectrum} shows the expected RIAF spectra. They comprise three components: a hard power law below $10^{12}$ Hz due to cyclo-synchrotron emission, a softer power law from IR to soft-X rays due to multiple IC upscattering by semi-relativistic electrons, and a bump peaking in soft $\gamma$-rays due to bremsstrahlung. With increasing $\dot{m}$, a conspicuous hardening of the IC component in the UV to X-ray range can be seen, besides the overall increase in the luminosity. This is a robust prediction of ADAF models, and reasonably representative of RIAFs in general (Yuan \& Narayan 2014). The significant differences in the RIAF spectra among the BL Lac subclasses have key consequences for their neutrino emission.


\section{{\si Neutrino emission induced by RIAFs}}
\label{sec:neutrino}

We now discuss the neutrino emission from BL Lacs, considering RIAFs as sources of external target photons for $p\gamma$ interactions with protons accelerated inside their jets. The basic formulation follows Tavecchio et al. (2014), of which the main points 1-3 are summarized below. Aspects newly considered for this work are described as point 4. All physical quantities as measured in the jet comoving frame are primed.


\begin{table} 
\centering
\begin{tabular}{c|ccccc}
\hline
Type  & $P_{\rm rad}$              & $P_{\rm jet}$                    &        $\dot{m}$    & $L^{\prime}_{\rm p}$   & $R_{\nu_\mu}  $  \\
                      &     erg s$^{-1}$                          &         erg s$^{-1}$                           &      $(10^{-3})$                     &    erg s$^{-1}$  & 7 yr\\
\hline   
{\si \bf LBL}                &$6.3 \cdot 10^{44} $ & $6.3 \cdot 10^{45}$ &  $5.7$ & $5 \cdot 10^{44}$ & $1$\\
\hline   
{\si \bf IBL}                 &$6.3 \cdot 10^{43}$ & $6.3 \cdot 10^{44}$ &    $0.5$ & $4.5 \cdot 10^{43}$ & $8\cdot 10^{-5}$\\
\hline    
{\si \bf HBL}               & $2.5 \cdot 10^{43} $ & $2.5 \cdot 10^{44}$ &  $0.3$  & $1.8 \cdot 10^{43}$ & $8 \cdot 10^{-7}$\\
\hline

\end{tabular}
\caption{Radiative power, jet power, normalized accretion rate, proton power and neutrino detection rate for the three subclasses of BL Lacs.
}
\label{tab:adaf}
\end{table}


{\si
1. Considering a region in the jet with radius $R_j = 10^{15}$ cm moving with bulk Lorentz factor $\Gamma_j=15$, accelerated protons are injected isotropically in the jet frame with luminosity $L'_{p}$, distributed in energy $E'_{p}$ as a power-law with a maximum cutoff:
\begin{equation}
L'_p(E'_p)=k_p E_p^{'-n}\text{exp}\left(-\frac{E'_p}{E'_{p, {\rm max}}}\right); \qquad  E'_p>E'_{p, {\rm min}}
\end{equation}
where, for definiteness, we set $E'_{p, {\rm max}}=10^{17}$ eV, $E'_{p, {\rm min}}=3\cdot 10^{11}$eV and $n=2$. Heavier nuclei are neglected.
}

{\si
2. The photomeson production efficiency $f_{p\gamma}(E'_p)$ is determined by the ratio between the dynamical timescale  $t'_{\text{dyn}} \approx R_j/c$ and $t'_{p\gamma}(E'_p)$, the energy loss timescale for protons via $p\gamma$ interactions.
}


{\si
3. The neutrino luminosity $L'_\nu$ in the jet frame is evaluated by (e.g., Murase et al. 2014):
\begin{equation}
E'_\nu L'_\nu(E'_\nu)\approx \frac{3}{8}f_{p\gamma}(E'_p)E'_p L'_p(E'_p); \qquad E'_\nu=0.05 E'_p.
\end{equation}
Using the Doppler factor of the emission region $\delta=[\Gamma_j(1-\beta_j \cos \theta)]^{-1}$, where $\beta_j =(1-1/\Gamma_j^2)^{1/2}$ and $\theta \approx 1/\Gamma_j$ is the viewing angle with respect to the jet axis, the luminosity of {\it muon} neutrinos $L_{\nu_\mu}$ in the observer frame is
\begin{equation}
E_\nu L_{\nu_\mu}(E_\nu) = \frac{1}{3}E'_\nu L'_\nu(E'_\nu )\delta^4; \qquad E_\nu = \delta E'_\nu.
\end{equation}
Note that the factor 1/3 accounts for equipartition among the flavors due to neutrino oscillations during propagation.
}

\begin{figure}
\hspace{0.5truecm}
  \includegraphics[width=.4\textwidth]{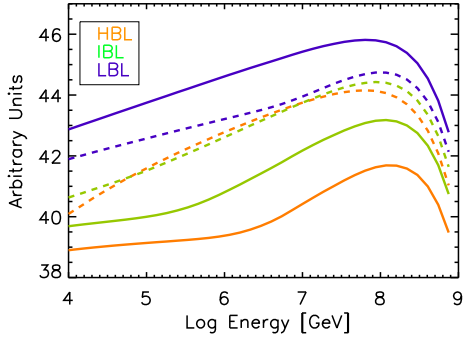} 
  \hspace{-0.4truecm}
  \caption{{\si Neutrino spectra due to $p\gamma$ interactions between protons with fixed $L^{\prime}_p=10^{45}$ erg s$^{-1}$
and external RIAF photons for the three subclasses of BL Lacs: LBL (solid blue), IBL (solid green), HBLs (solid orange). Contributions from internal photons as $p\gamma$ targets are also shown (dashed blue, green and orange for LBL, IBL, HBL, respectively).}}
\label{fig:nuspectrum}
\end{figure}


4. To evaluate $t'_{p\gamma}(E'_p)$ for this work, we account for both internal synchrotron photons from electrons accelerated in the jet, and external photons from the RIAF. For the internal photons, we utilize the SED models for the observed non-thermal emission described in \S 2, assume that it originates co-spatially with the protons and isotropically in the jet frame, and convert the SEDs into photon density in the jet frame using $\delta$. For external photons from the RIAF, we utilize the models of \S 3 and make the simplifying assumption that in the jet frame, they are nearly isotropic and uniform with energy density $\Gamma_j^2 /3$ times its value in the BH frame, and that the jet emission region is at distance $d=10^{16}$ cm from the BH.

Regarding the latter, we note that in many analytic descriptions of RIAFs, UV-X-ray photons are expected to emerge primarily from within a few $r_S$ from the BH. Such photons would enter the jet region mostly from behind, appearing substantially {\it debeamed} in the jet frame. However, detailed numerical models of RIAF-jet systems show that the jet can be surrounded by a funnel formed by a relatively dense wind (e.g. Sadowski et al. 2013), which could scatter and isotropise a fair fraction of the RIAF photons before they enter the jet. Moreover, Ryan et al. (2017) show that the funnel contains hot electrons that contribute appreciably to UV-X-ray emission out to $\sim$15 $r_S$ (see also Ryan et al. 2018, Chael et al. 2018 and Nakamura et al. 2018 for the specific case of M87). Thus, external UV-X-ray photons can impinge into the jet region up to angles $\approx \pi/2$ from the jet axis. For such geometries, the energy density of external photons in the jet frame is amplified by a factor $f\Gamma_j^2$, where $f\simeq 1/3$ for an isotropic distribution, adopted here for simplicity.

{\si
The resulting neutrino spectra for each BL Lac subclass are compared in Fig. \ref{fig:nuspectrum}, which also shows the contributions from external RIAF and internal photons separately. To highlight the effect of the different RIAF spectra, here $L^{\prime}_p=10^{45}$ erg s$^{-1}$ has been fixed. Most notably, the neutrino luminosity of LBLs at $E_\nu \sim$ 0.1-1 PeV is $\sim4$ orders of magnitude larger than that of HBLs, primarily due to the significant difference in the density of external RIAF photons in the soft X-ray range, which serve as the main $p\gamma$ targets for protons with $E_p \sim$ 2-20 PeV. We also see that while internal photons are the most prevalent $p\gamma$ targets in HBLs, external RIAF photons become relatively more important in IBLs, and completely dominate in LBLs.
}

\begin{figure}
\vspace*{-0.5truecm}
\hspace*{-1.4truecm}
\centering
 \includegraphics[width=.4\textwidth]{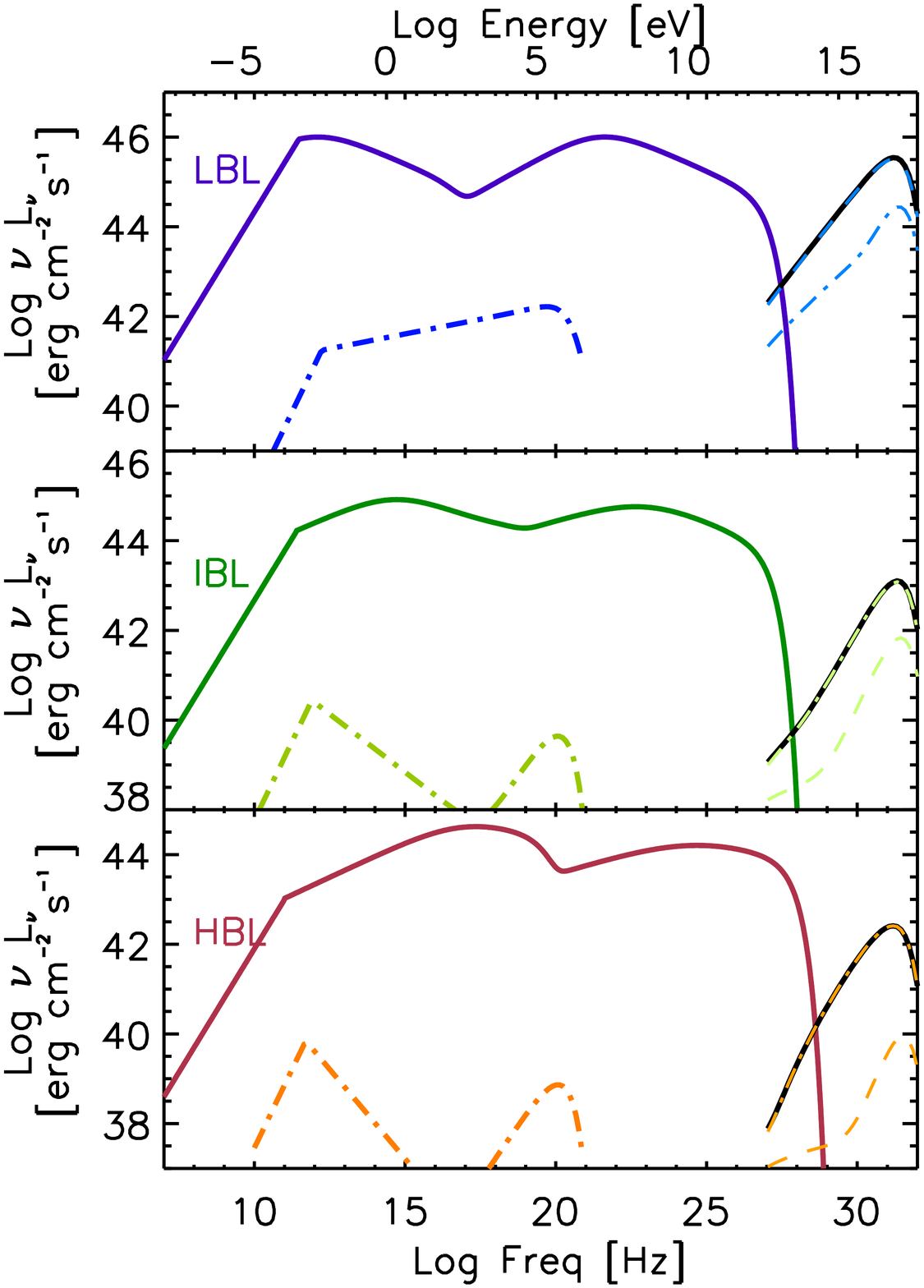} 
  \caption{{\si SEDs for the three subclasses of BL Lacs: LBL (top), IBL (middle), HBL (bottom), showing the electromagnetic components from the jet (solid colored) and RIAF (dot-dashed), and the neutrino components due to internal photons (dotted), external RIAF photons (dashed), and their sum (solid black). Note the different scales for luminosity between the panels.}}
  \label{fig:sed1}
\end{figure}

{\si
More realistically, $L'_p$ is likely linked to $P_{\rm jet}$ and is expected to vary among the BL Lac subclasses. An important test case is the BL Lac TXS 0506+056, potentially associated with IceCube-170922A, a $\sim$300 TeV neutrino (Aartsen et al. 2018). While TXS 0506+056 may be classifiable as an IBL from the observed $\nu_S$ alone, its observed luminosity is more representative of an LBL, especially in terms of our SED classification discussed in \S 2. We assume that TXS 0506+056 is a typical LBL, emitting neutrinos according to our model that includes external RIAF photons. With the measured redshift of $z=0.3365 \pm 0.0010$ (Paiano et al. 2018) and the IceCube effective area appropriate for the declination of TXS 0506+056 (Aartsen et al. 2008), its neutrino flux must be high enough to result in at least one $\nu_\mu$ detection during 7 years of IceCube observations in the energy range 60 TeV - 10 PeV, roughly corresponding to uncertainty for IceCube-170922A. This translates into a constraint on $L'_p$ for LBLs. The values for IBLs and HBLs follow by assuming $L'_p \propto P_{\rm jet}$. With these values of $L'_p$ for the different BL Lac subclasses, their neutrino spectra can be predicted as shown in Fig. \ref{fig:sed1}, together with the corresponding SEDs of the electromagnetic emission from the jet and RIAF. Tab. \ref{tab:adaf} lists the values of $L'_p$ and $R_{\nu_\mu}$, the neutrino detections expected in 7 years.
}

{\si
Compared to the case assuming constant $L'_p$, the differences between LBLs and the other, less luminous subclasses is naturally magnified. As above, RIAFs play a significant role only for LBLs. In this scheme, only LBLs may be sufficiently powerful neutrino emitters to be observationally relevant. These inferences for the RIAF model are particularly interesting in view of the fact that the LBL TXS 0506+056 is likely the first identified source of high-energy neutrinos, while HBLs such as Mkn 421 and Mkn 501 are yet to be detected by IceCube, despite being conspicuous, nearby $\gamma$-ray emitters with some predictions of detectability (e.g. Petropoulou et al. 2015). Although the  statistics is currently limited, stronger tests of this picture through further observations are anticipated.}

\section{Discussion}


{\si
We have conducted a first study of the role of RIAFs as sources of external target photons for $p\gamma$ neutrino production in BL Lacs, finding that they can be particularly relevant for the subclass of LBLs, but less so for IBLs or HBLs. These results have interesting implications for interpreting the potential association of IceCube-170922A with the LBL TXS 0506+056, and the non-detections by IceCube so far of HBLs such as Mkn 421 and Mkn 501.
}

{\si
As an exploratory step, many simplifying assumptions were made concerning various aspects, which deserve more detailed and comprehensive considerations in the future. We have assumed rudimentary scaling relations between $P_{\rm rad}$, $P_{\rm jet}$ and $\dot{M}$, and fixed quantities such as $M_{\rm BH} $ for simplicity. A more realistic study obviously needs to account for the distribution and scatter of these variables. Although our description of the non-thermal electromagnetic emission was entirely phenomenological, more physical modelling is warranted, including the potential effects of EC emission induced by RIAFs, hadronic emission components triggered by $p\gamma$ interactions, etc.

The simple ADAF prescription of Mahadevan (1997) that we employed can be updated with more advanced RIAF models (Yuan \& Narayan 2014). Since the RIAF is geometrically thick, with different spatial dependences for each of its spectral components, accurate evaluations require a more proper treatment of the spatial and angular distribution of the RIAF photons impinging into the jet, which can also be affected by electron scattering in the jet vicinity. Such calculations may reveal non-trivial beaming patterns for both the EC and neutrino emission, with potentially important observational implications (see relevant discussion in Ansoldi et al. 2018).}

Alternative scenarios have been proposed for neutrino emission from BL Lacs. Models in which the $\gamma$-rays are dominated by hadronic processes may allow more luminous neutrino emission than conventional expectations, but generally at the expense of a high value for $L'_p$ that may not be realistic except in certain cases (e.g. Cerruti et al. 2015, Petropoulou \& Dermer 2016). Spine-sheath scenarios were proposed by Tavecchio et al. (2014, 2015) and Righi et al. (2017), where the jet consists of a faster spine structure enveloped by a slower sheath structure (e.g. Ghisellini et al. 2005), such that synchrotron photons from the sheath can serve effectively as external photons for the spine, enhancing the neutrino yield compared to cases with only internal photons as $p\gamma$ targets. This scenario can provide a self-consistent explanation for TXS 0506+056 and IceCube-170922A (Ansoldi et al. 2018), and would generally predict that all subclasses of BL Lacs can be efficient neutrino emitters.
In contrast, the RIAF scenario naturally favours only LBLs as significant neutrino sources, potentially in accord with the current observational status.
Another advantage of the scenario is that the spectrum of target photons is uniquely prescribed by the RIAF model, unlike the spine-sheath scenario for which the spectrum of the sheath radiation is not well defined a-priori. Further observations by IceCube combined with multiwavelength follow-up efforts should provide definitive discrimination among different models for neutrino emission from BL Lacs.

Despite their expectation as promising neutrino sources, no FSRQ has been clearly detected to date. One possibility is that their jet composition is predominantly electron-positron so that very few protons are accelerated therein, while that of BL Lacs is predominantly electron-proton, a hypothesis for which there is some observational support (e.g. Hardcastle 2018). Low-power radio galaxies, the parent population of BL Lacs with their jets oriented away from the observer, may also be potentially interesting neutrino sources, although the present model tailored to BL Lacs cannot be readily applied to such objects. Further discussion of FSRQs and radio galaxies is beyond the current scope and deferred to future work.


\section*{Acknowledgments}
We thank the referee for valuable and constructive comments. CR and FT acknowledge support from the INAF CTA--SKA grant "Probing particle acceleration and $\gamma$-ray propagation with CTA and its precursors". SI thanks support from JSPS KAKENHI Grant Number JP17K05460.

%


\begin{thebibliography}{}

\bibitem[\protect\citeauthoryear{Aartsen et al.}{2013}]{2013PhRvL.111b1103A} Aartsen M.~G., et al., 2013, PhRvL, 111, 021103 
\bibitem[\protect\citeauthoryear{Aartsen et al.}{2017}]{} Aartsen M.~G., et al., 2017, ApJ, 835, 151 
\bibitem[\protect\citeauthoryear{Aartsen et al.}{2018}]{} Aartsen M.~G., et al., 2018, Science, 361, eaat1378 
\bibitem[\protect\citeauthoryear{Ackermann et al.}{2015}]{2015ApJ...810...14A} Ackermann M., et al., 2015, ApJ, 810, 14 
\bibitem[\protect\citeauthoryear{Ahlers \& Halzen}{2015}]{2015RPPh...78l6901A} Ahlers M., Halzen F., 2015, RPPh, 78, 126901 
\bibitem[\protect\citeauthoryear{Ansoldi et al.}{2018}]{2018ApJ...863L..10A} Ansoldi S., et al., (The MAGIC Coll.) 2018, ApJ, 863, L10 
\bibitem[\protect\citeauthoryear{Celotti \& Ghisellini}{2008}]{2008MNRAS.385..283C} Celotti A., Ghisellini G., 2008, MNRAS, 385, 283 
\bibitem[\protect\citeauthoryear{Cerruti et al.}{2015}]{2015MNRAS.448..910C} Cerruti M., Zech A., Boisson C, Inoue S., 2015, MNRAS, 448, 910 
\bibitem[\protect\citeauthoryear{Chael, Narayan, \& Johnson}{2018}]{2018arXiv181001983C} Chael A., Narayan R., Johnson M.~D., 2018, MNRAS, submitted (arXiv:1810.01983) 
\bibitem[\protect\citeauthoryear{Dermer \& Menon}{2009}]{2009herb.book.....D} Dermer C.~D., Menon G., 2009, High Energy Radiation from Black Holes: Gamma Rays, Cosmic Rays, and Neutrinos, Princeton University Press 
\bibitem[\protect\citeauthoryear{Fossati et al.}{1998}]{1998MNRAS.299..433F} Fossati G., Maraschi L., Celotti A., Comastri A., Ghisellini G., 1998, MNRAS, 299, 433 
\bibitem[\protect\citeauthoryear{Ghisellini et al.}{2017}]{2017MNRAS.469..255G} Ghisellini G., Righi C., Costamante L., Tavecchio F., 2017, MNRAS, 469, 255 (G17)
\bibitem[\protect\citeauthoryear{Ghisellini, Maraschi, \& Tavecchio}{2009}]{2009MNRAS.396L.105G} Ghisellini G., Maraschi L., Tavecchio F., 2009, MNRAS, 396, L105 
\bibitem[\protect\citeauthoryear{Ghisellini, Tavecchio, \& Chiaberge}{2005}]{2005A&A...432..401G} Ghisellini G., Tavecchio F., Chiaberge M., 2005, A\&A, 432, 401 
\bibitem[\protect\citeauthoryear{Ghisellini et al.}{2010}]{2010MNRAS.402..497G} Ghisellini G., Tavecchio F., Foschini L., Ghirlanda G., Maraschi L., Celotti A., 2010, MNRAS, 402, 497 
\bibitem[\protect\citeauthoryear{Ghisellini et al.}{2014}]{2014Natur.515..376G} Ghisellini G., Tavecchio F., Maraschi L., Celotti A., Sbarrato T., 2014, Nature, 515, 376 
\bibitem[\protect\citeauthoryear{Hardcastle et al.}{2018}]{2018arXiv181107943H} Hardcastle M.~J., et al., 2018, arXiv, arXiv:1811.07943 

\bibitem[\protect\citeauthoryear{Kadler et al.}{2016}]{2016NatPh..12..807K} Kadler M., et al., 2016, NatPh, 12, 807
\bibitem[\protect\citeauthoryear{Kowalski}{2015}]{2015JPhCS.632a2039K} Kowalski M., 2015, JPhCS, 632, 012039 
\bibitem[\protect\citeauthoryear{Madejski \& Sikora}{2016}]{2016ARA&A..54..725M} Madejski G., Sikora M., 2016, ARAA, 54, 725 
\bibitem[\protect\citeauthoryear{Mahadevan}{1997}]{1997ApJ...477..585M} Mahadevan R., 1997, ApJ, 477, 585 
\bibitem[\protect\citeauthoryear{Mannheim}{1993}]{1993A&A...269...67M} Mannheim K., 1993, A\&A, 269, 67 
\bibitem[\protect\citeauthoryear{Mannheim}{1995}]{1995APh.....3..295M} Mannheim K., 1995, Astropart. Phys., 3, 295 
\bibitem[\protect\citeauthoryear{Meszaros}{2017}]{2017ARNPS..67...45M} Meszaros P., 2017, ARNPS, 67, 45 
\bibitem[\protect\citeauthoryear{Murase \& Waxman}{2016}]{2016PhRvD..94j3006M} Murase K., Waxman E., 2016, PhRvD, 94, 103006 
\bibitem[\protect\citeauthoryear{Murase, Inoue, \& Dermer}{2014}]{2014PhRvD..90b3007M} Murase K., Inoue Y., Dermer C.~D., 2014, PhRvD, 90, 023007 
\bibitem[\protect\citeauthoryear{Nakamura et al.}{2018}]{2018arXiv181009963N} Nakamura M., et al., 2018, ApJ, in press (arXiv:1810.09963) 
\bibitem[\protect\citeauthoryear{Narayan \& Yi}{1994}]{1994ApJ...428L..13N} Narayan R., Yi I., 1995, ApJ, 452, 710 
\bibitem[\protect\citeauthoryear{Nemmen et al.}{2012}]{2012Sci...338.1445N} Nemmen R.~S., Georganopoulos M., Guiriec S., Meyer E.~T., Gehrels N., Sambruna R.~M., 2012, Sci, 338, 1445
\bibitem[\protect\citeauthoryear{Paiano et al.}{2018}]{2018ApJ...854L..32P} Paiano S., Falomo R., Treves A., Scarpa R., 2018, ApJ, 854, L32
\bibitem[\protect\citeauthoryear{Petropoulou \& Dermer}{2016}]{2016ApJ...825L..11P} Petropoulou M., Dermer C.~D., 2016, ApJ, 825, L11 
\bibitem[\protect\citeauthoryear{Petropoulou et al.}{2015}]{2015ICRC...34.1125P} Petropoulou M., Dimitrakoudis S., Padovani P., Resconi E., Giommi P., Mastichiadis A., 2015, ICRC, 34, 1125 
\bibitem[\protect\citeauthoryear{Righi, Tavecchio, \& Guetta}{2017}]{2017A&A...598A..36R} Righi C., Tavecchio F., Guetta D., 2017, A\&A, 598, A36
\bibitem[\protect\citeauthoryear{Ryan et al.}{2017}]{2017ApJ...844L..24R} Ryan B.~R., Ressler S.~M., Dolence J.~C., Tchekhovskoy A., Gammie C., Quataert E., 2017, ApJ, 844, L24
\bibitem[\protect\citeauthoryear{Ryan et al.}{2018}]{2018ApJ...864..126R} Ryan B.~R., Ressler S.~M., Dolence J.~C., Gammie C., Quataert E., 2018, ApJ, 864, 126 
\bibitem[\protect\citeauthoryear{S{\c a}dowski et al.}{2013}]{2013MNRAS.436.3856S} S{\c a}dowski A., Narayan R., Penna R., Zhu Y., 2013, MNRAS, 436, 3856 
\bibitem[\protect\citeauthoryear{Shakura \& Sunyaev}{1973}]{1973A&A....24..337S} Shakura N.~I., Sunyaev R.~A., 1973, A\&A, 24, 337 
\bibitem[\protect\citeauthoryear{Tavecchio \& Ghisellini}{2015}]{2015MNRAS.451.1502T} Tavecchio F., Ghisellini G., 2015, MNRAS, 451, 1502 
\bibitem[\protect\citeauthoryear{Tavecchio, Ghisellini, \& Guetta}{2014}]{2014ApJ...793L..18T} Tavecchio F., Ghisellini G., Guetta D., 2014, ApJ, 793, L18 
\bibitem[\protect\citeauthoryear{Tchekhovskoy, Narayan, \& McKinney}{2011}]{2011MNRAS.418L..79T} Tchekhovskoy A., Narayan R., McKinney J.~C., 2011, MNRAS, 418, L79 
\bibitem[\protect\citeauthoryear{Yuan \& Narayan}{2014}]{2014ARA&A..52..529Y} Yuan F., Narayan R., 2014, ARAA, 52, 529 
\end{thebibliography}
\end{document}